\documentclass[intlimits,twoside,a4paper]{article}
\usepackage[cp1251]{inputenc} 
\usepackage[eqsecnum]{cmpj3}
\issue{2024}{27}{1}{13605}
\doinumber{10.5488/CMP.27.13605}

\title[Finite size effects]
      {Finite size effects and optimization of the calculation
        of the surface tension in surfactant
    mixtures at liquid/vapour interfaces}

\author[C. J. Guzman-Valencia, J. Toriz-Salinas, H. Espinosa-Jimenez, A. B.
Salazar-Arriaga, J. L. L\'{o}pez-Cervantes, H. Dominguez]
{C. J. Guzman-Valencia\refaddr{label1}, J. Toriz-Salinas\refaddr{label1},
	H. Espinosa-Jimenez\refaddr{label1}, A.~B.~Salazar-Arriaga\refaddr{label1},
	J. L. L\'{o}pez-Cervantes\refaddr{label2}, 
	H. Dominguez\orcid{0000-0001-6126-9300}\refaddr{label1}\thanks{Corresponding author: \email{hectordc@unam.mx}.}}

\addresses{
	\addr{label1}Instituto de Investigaciones en Materiales, 
	Universidad Nacional Aut\'{o}noma de M\'{e}xico, UNAM Cd. Mx. 04510, M\'{e}xico.
	\addr{label2}Facultad de Qu\'{i}mica, Universidad Nacional Aut\'{o}noma de M\'{e}xico,
	UNAM Cd. Mx. 04510, M\'{e}xico}


\date{Received January 04, 2024, in final form February 21, 2024}

\begin{document}

\maketitle

\begin{abstract}

The surface tension of monolayers with mixtures of anionic and nonionic
surfactant  at the liquid/vapour interface is studied.
Previous works have observed that calculations of the surface tension
of simple fluids show artificial oscillations for small interfacial areas,
indicating that the surface tension data fluctuate due to the finite size
effects and periodic boundary conditions. In the case of simulations
of monolayers composed of surfactant mixtures, the surface tension
not only oscillates for small areas but can also give non-physical
data, such as negative values. Analysis of the monolayers
with different surfactant mixtures, ionic (DTAB, CTAB, SDS) and
nonionic (SB3-12), was done for density profiles, parameters of order and pair
correlation functions for small and large box areas
and all of them present similar behaviour.
The fluctuations and the non-physical values of the surface tension
are corrected when boxes with large interfacial areas are considered.
The results indicate that in order to obtain reliable values of the
surface tension, in computer simulations, it is important to choose not
only the correct force field but also the appropriate size of the
simulation box.

\printkeywords
%
%
%\pacs 68.03.Cd; 68.18.Fg; 68.03.Hj; 82.70.Uv
\end{abstract}

%\newpage

\section{Introduction}

 Surfactant molecules have been extensively studied in different
interfacial problems not only for their scientific interest but also for
their numerous industrial applications. Therefore, several experimental techniques
have been used to investigate their properties \cite{1,2,3,3a,3b,3c},
in particular the surface tension \cite{4,5,5a,5b,5c,5d}. Although,
most of the studies are conducted for one type of surfactant,
a lot of commercial products consist of a mixture of molecules
which present richer properties. For example, anionic
surfactants are usually mixed with non-anionic ones in products such as shampoos,
dish washing liquids, washing powders, whereas mixtures of cationic with
non-ionic surfactants are good for disinfectants, cleaners and sanitizers.
On the other hand, the capability of surfactants to be
adsorbed at liquid interfaces is an important property and it is widely
used for many technological processes, such as detergents, foam, and emulsion
stabilizers. In fact, adsorption of surfactant solutions at
liquid interfaces has been investigated in many papers using mixtures of
non-ionic with anionic or cationic surfactants \cite{6,7,7a}.

  In particular, synergistic effects of mixtures of anionic and
cationic surfactants are relevant in several industrial applications.
In experiments, the synergy can be determined
by measuring the surface tension as a function of concentration
at different surfactant ratios, which is time consuming. Therefore, the
combination of computer methods and experiments will reduce
the experimental times and provide microscopic insight into the mechanism
of the synergistic effect.

Since computer simulations are relevant to understanding the interfacial problems, they
 have appeared as an alternative to the study of such
complex systems. From the computer simulations it is possible to
obtain more data about thermodynamic properties, such
as the surface tension, that are not easy
to collect from actual experiments. 
However, to achieve this goal, it is
imperative to have good computer methods to gain reliable results, i.e.,
not only the force field is important but also the conditions to carry out
the simulations. For instance, the calculation of the
surface tension can depend on the  number of molecules and
the cutoff values used in the simulations \cite{8,9,10,11}.
Therefore, some papers have investigated the effects
of truncated potentials \cite{8}, the inclusion of long-range
corrections \cite{12}, the finite size and the periodic boundary conditions
in Lennard-Jones fluids at the liquid/vapour interface. 
Moreover, some authors have shown that the surface tension data present 
an oscillatory behaviour \cite{12} and they concluded that reasonable values
of the surface tension can be obtained with large interfacial
areas \cite{12} and cutoff values of about 2.5 nm \cite{13}.

 In the present work we extend previous studies and
calculate the surface tension of systems composed of monolayers
with one type and mixtures of surfactant molecules at the liquid/vapour
interface. We show that choosing incorrect simulations boxes can give
erroneous surface tension results, including values that are not physically
acceptable as negative values. The errors can be corrected if boxes with
large interfacial areas are taken. Simulations were carried out
by using different ionic and nonionic surfactants.

\begin{figure}[h]
%\vspace{1cm}
\begin{center}
\includegraphics[width=3.2in]{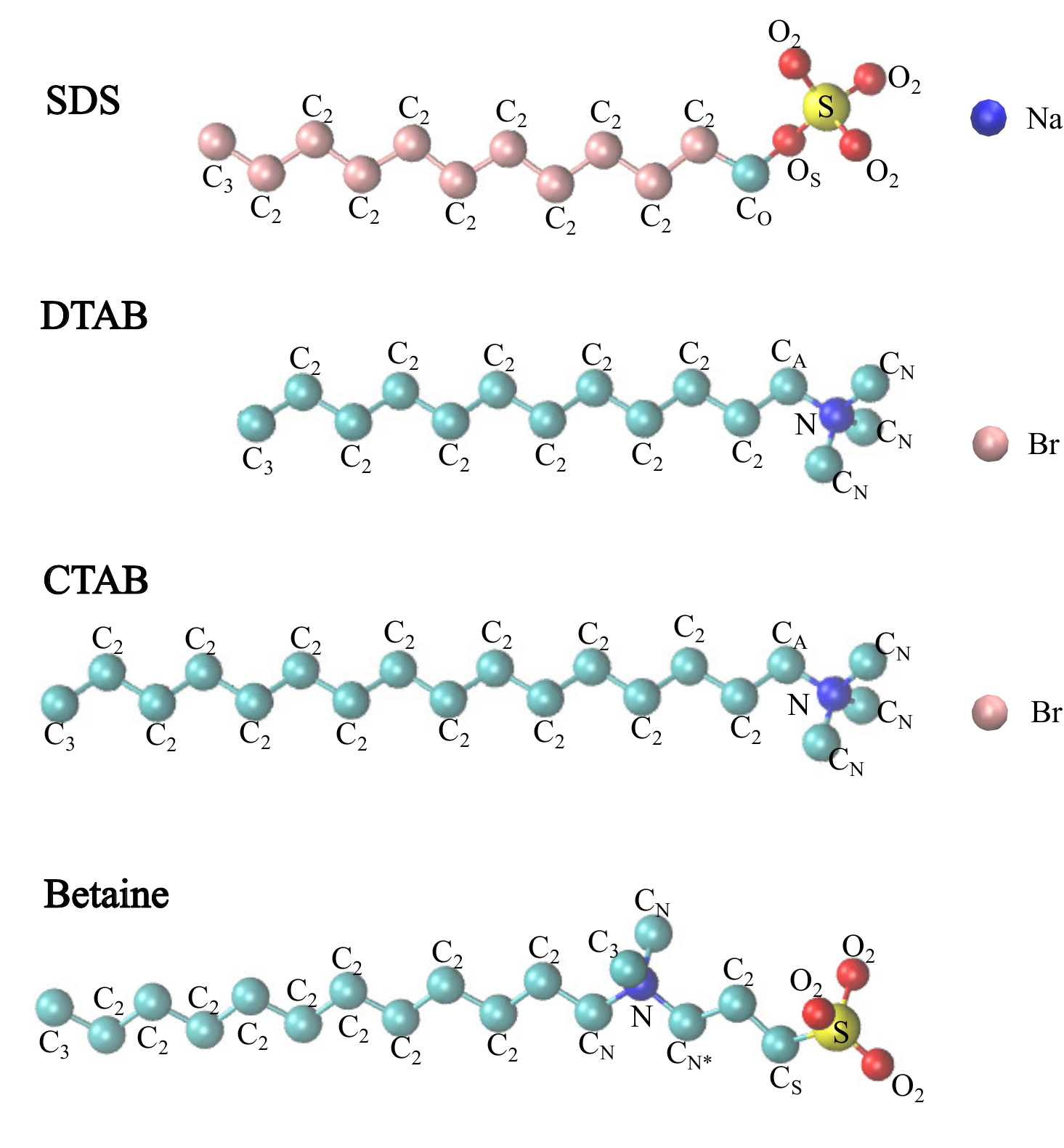}
\caption{(Colour online) Molecular structure for each surfactant used in the
  simulations within the united atom model.}
  \label{fig1} 
\end{center}
\end{figure}

\section{Computational model}

 Different surfactant molecules were simulated, the sodium dodecyl
sulfate (SDS), the dodecyltrimethylammonium bromide (DTAB), the
cetyltrimethylammonium bromide (CTAB) and the zwitterionic lauryl
sulfobetaine (SB-3), called Betaine in figure~\ref{fig1}. The first one is anionic,
the next two are cationic  and the last one is nonionic
surfactant, respectively (see figure \ref{fig1}).
In all cases the atoms of the surfactant headgroups were explicitly simulated 
whereas the CH$_n$ groups in the hydrocarbon tails were modelled using
the united atom model  (see figure \ref{fig1}). Counterions, Na$^+$ and Br$^-$,
were also introduced into the water phase
to neutralize the SDS and DTAB (and CTAB) systems. For water,
the TIP4P/$\epsilon$ model was used \cite{14} and all the
unlike interactions were handled with the Lorentz-Berthelot rules \cite{15}.

 Initially, simulations of monolayers with one type of surfactant
were carried out to validate the force field parameters of all the
surfactants. For these simulations four monolayers were prepared, each one with
45 SDS, 45 DTAB, 45 CTAB and 45 Betaine molecules, which were placed
at the liquid/vapour interface 
in a rectangular box with dimensions $L_x = L_y = 5$~nm and
$L_z = 15$~nm. Then, the surface tensions were calculated and
compared with the experimental data. In some cases, the Lennard-Jones (LJ)
parameters of the surfactants were reparametrized to obtain a better
agreement between the simulations and the experiments (see
the next section). With the LJ parameters obtained from previous
calculations other series of simulations were executed
for monolayers prepared with mixtures of CTAB/SDS, CTAB/DTAB and
CTAB/Betaine surfactants at different compositions.

 All simulations were  carried out using the Gromacs
software \cite{15a} in the canonical (NVT) ensemble at
constant temperature, $T = 298$~K, using the Nos\'{e}-Hoover thermostat \cite{16}
with a relaxation time constant of $\tau_{T} = 1$~ps.
Electrostatic interactions were handled
with the particle mesh Ewald method \cite{17} and the short range interactions
were cutoff at 2.5 nm as suggested in reference \cite{13}.
Bond lengths were constrained using the Lincs algorithm \cite{18}.
Then, simulations were carried out up to 50 ns after 10 ns of equilibration
with a time step of $\rd t = 0.002$~ps, and collecting data, over the last 20 ns,
were taken for data analysis. Equilibration of the systems
was monitored with the configurational energy and the calculated structural
properties (see the results section), which did not change significantly
along the simulation time.

\begin{table}[ht]
	\caption{\label{tab1} \small{SDS force field parameters, see
			figure \ref{fig1} for labels.}}
	\small
	\begin{center}
		%\vspace{.5cm}
		\begin{tabular}{|c|c|c|c|}
			\hline
			Site  & $q(e)$ & $\sigma_{LJ}$(\AA) & $\epsilon_{LJ}$(KJ/mol)\\
			\hline
			C$_3$       & 0.0000 & 3.3997 & 0.45773 \\
			C$_2$       & 0.0000 & 3.3997 & 0.27000 \\
			C$_O$       & 0.1370 & 3.3997 & 0.35982 \\
			O$_S$  & $-0.4590$ & 3.0000 & 0.71128 \\
			O$_2$   & $-0.6540$ & 2.9599 & 0.87864 \\
			S           & 1.2829  & 3.5636 & 1.04600 \\
			Na          & 1.0000  & 3.3284 & 0.01159 \\
			\hline
		\end{tabular}
	\end{center}
\end{table}

\begin{table}[h]
	\caption{\label{tab2} \small{CTAB/DTAB force field parameters, see
			figure \ref{fig1} for labels.}}
	\small
	\begin{center}
		%\vspace{.5cm}
		\begin{tabular}{|c|c|c|c|}
			\hline
			Site  & $q(e)$ & $\sigma_{LJ}$(\AA) & $\epsilon_{LJ}$(KJ/mol) \\
			\hline
			C$_3$  & 0.0000 & 3.3997 & 0.45773\\
			C$_2$  & 0.0000 & 3.3997 & 0.31800\\
			C$_A$ & 0.2167 & 3.3997 & 0.35982\\
			C$_N$  & 0.1701 & 3.3996 & 0.45773\\
			N      & 0.2730 & 3.2500 & 0.71128\\
			Br     & $-1.0000$ & 3.95556 & 1.3388\\
			\hline
		\end{tabular}
	\end{center}
\end{table}

\section{Results}
\subsection{Monolayers with a single type of surfactant}

 As stated above, initial simulations of monolayers
with a single type of surfactant were performed to reparametrize the
surfactant LJ parameters and obtain reasonable values of the
surface tensions. Reparametrization was carried out by 
the three step systematic parametrization procedure, 3SSPP \cite{19},
where all the $\epsilon$ and $\sigma$ LJ
parameters were scaled from their original values until
the surface tension measurements were in good agreement with
the experiments, typically within 5\% error.
The procedure was conducted for the CTAB and SDS monolayers whereas
for the Betaine monolayer the parameters were taken from reference \cite{19a}.
In the molecular dynamics, the surface tension was calculated using
the expression,

\begin{equation}
  \gamma = L_z \left[\langle P_{zz} \rangle - \frac{1}{2}
\left(\langle P_{xx} \rangle + \langle P_{yy} \rangle \right)\right],
\label{eq_3_1}
\end{equation}
 where $L_z$ is the length of the simulation box and
$P_{ii}$ are the components of the pressure tensor,
$P_{ii}$ = $(1/V)(E_k - VIR)$ with $E_k$ the kinetic energy,
$V$ the volume and $VIR$ the virial expression \cite{15a},

\begin{equation}
  VIR = - \sum_{i<j} \bf{r_{ij}} \otimes \bf{F_{ij}}.
\end{equation}
 $\bf{r_{ij}}$ is the position vector between atom $i$ and atom $j$, and
$\bf{F_{ij}}$ is the force vector between those atoms.
Equation~\eqref{eq_3_1} is appropriate to calculate the surface tension
with ``planar'' interfaces, i.e., without pronounced curvatures.

\begin{table}[h]
	\caption{\label{tab3} \small{Betaine force field parameters, see figure \ref{fig1} for labels.}}
	\small
	\begin{center}
		%\vspace{.5cm}
		\begin{tabular}{|c|c|c|c|}
			\hline
			Site  & $q(e)$ & $\sigma_{LJ}$(\AA) & $\epsilon_{LJ}$(KJ/mol)\\
			\hline
			C$_3$ & 0.0000& 3.8709 & 0.68096\\
			C$_2$ & 0.0000 & 3.8707 & 0.29988\\
			C$_A$ & 0.1891 & 3.8709 & 0.29988\\
			C$_N*$ & 0.1311 & 3.8707 & 0.29988\\
			C$_S$ & 0.0842 & 3.8709 & 0.29988\\
			C$_N$ & 0.1902 & 3.8709 & 0.68096\\
			N     & 0.2407 & 3.8707 & 0.56015\\
			O$_3$ & $-0.7184$ & 2.9304 & 0.56015\\
			S     & 1.1297 & 3.5145 & 0.82373\\
			\hline
		\end{tabular}
	\end{center}
\end{table}

 In our case we have two interfaces, water/air at one side of the box
and water/surfactant/air at the other. Then,
from the gromacs software the surface tension is obtained,
\begin{equation}
\gamma = \gamma_{\rm water/air} + \gamma_{\rm water/surfactant/air}\,.
\end{equation}  
 By knowing the $\gamma_{\rm water/air}$,
the $\gamma_{\rm water/surfactant/air}$ can be estimated.
Therefore, simulations for the water/air interface were carried out
and a value of 69 mN/m was found, in good agreement with the
experimental data (72 mN/m).
The new surfactant Lennard-Jones parameters and the surface tensions for
the monolayers with single types of surfactants are given in
tables \ref{tab1}--\ref{tab4}.

\begin{table}[h]
	\caption{\label{tab4} \small{Surface tension data of all the surfactants, experiments, calculated in this work.}}
	\small
	\begin{center}
		%\vspace{.5cm}
		\begin{tabular}{|c|c|c|}
			\hline
			Surfactant  & $\gamma$ (mN/m) exp & $\gamma$ (mN/m) This work\\
			\hline
			CTAB   &36.6 \cite{Adamczyk}, 33.8 \cite{Kuperkar}  & 32.4  \\
			SDS  &34.1 \cite{Addison}, 38.2 \cite{Mysels}  & 37.5 \\
			Betaine   & 37.9 \cite{Xiao}  &  36.6 \\
			DTAB & 39.2\cite{Yamanaka}, 36.9 \cite{Shah}  & 36.5 \\
			\hline
		\end{tabular}
	\end{center}
\end{table}

\subsection{Monolayers with surfactant mixtures}

\begin{figure}[h]
	\vspace{1cm}
	\begin{center}
		\includegraphics[scale=0.5]{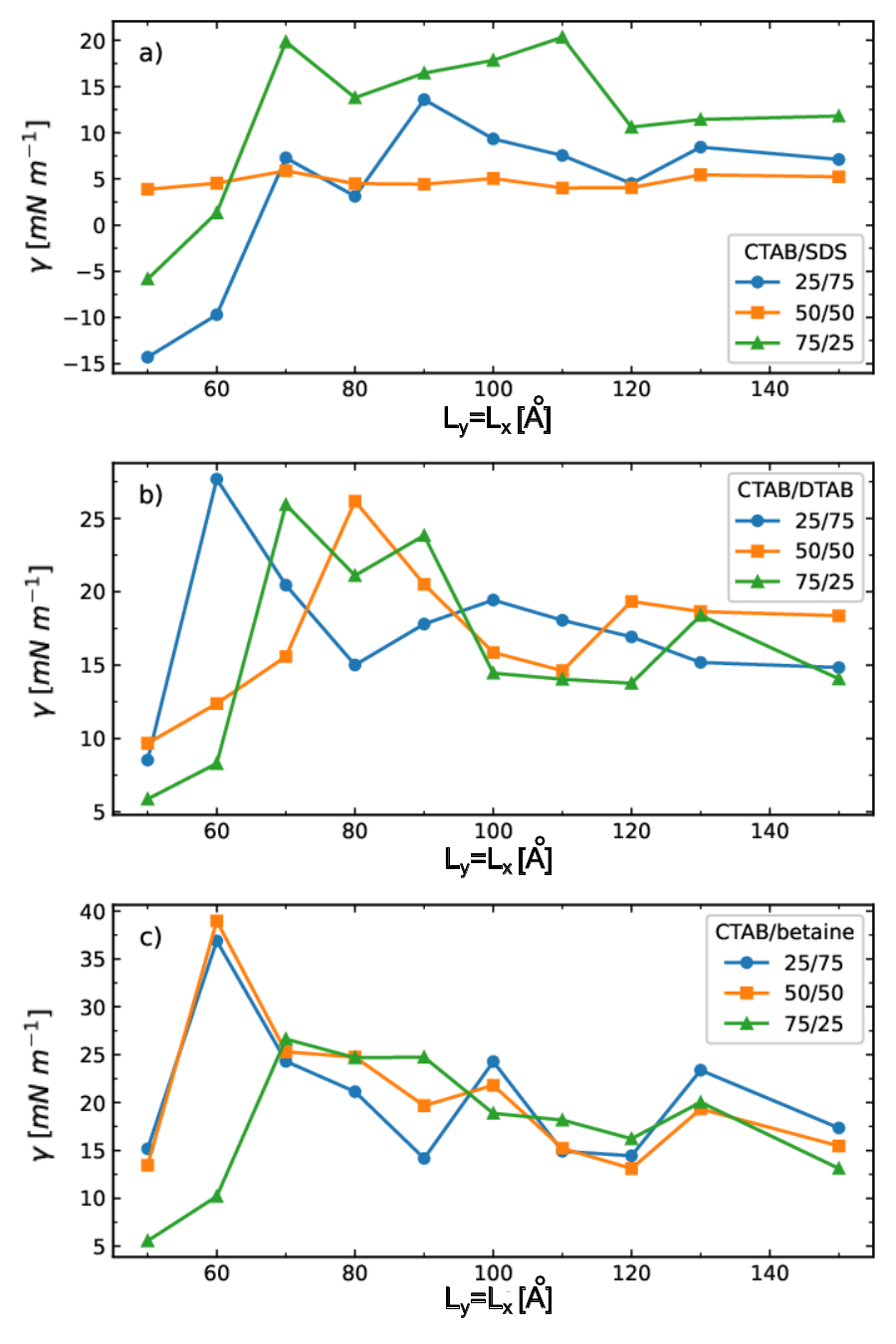}
		\caption{(Colour online) Surface tension data for the monolayers with mixtures of
			a) CTAB/SDS, b)~CTAB/DTAB and CTAB/Betaine at different surfactant
			compositions as function of the simulation box lengths.}
		\label{fig2} 
	\end{center}
\end{figure}

\begin{figure}[h]
%	\vspace{1cm}
	\begin{center}
		\includegraphics[width=3.5in]{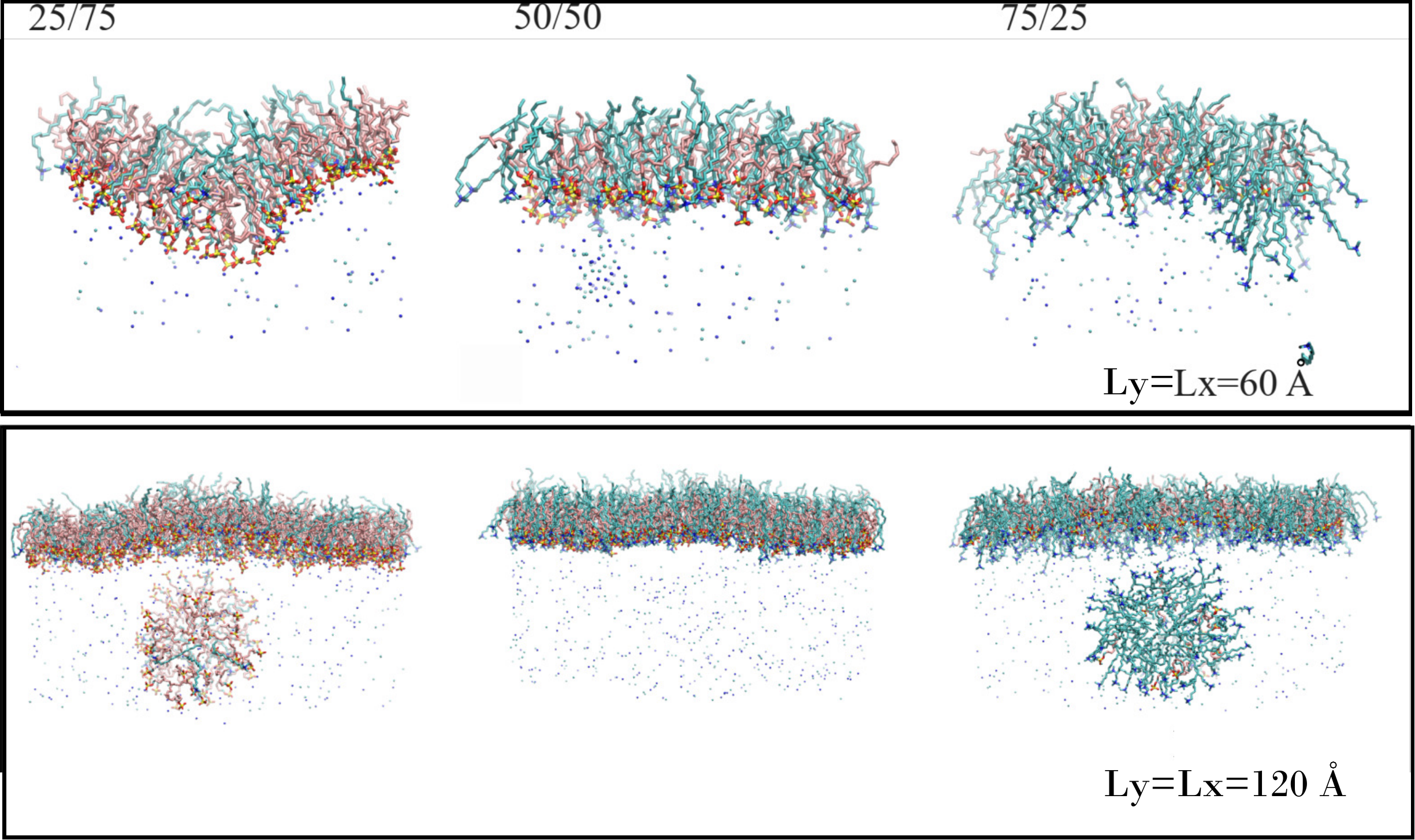}
		\caption{(Colour online) Snapshots of the last configurations of the monolayers
			with mixtures of CTAB/SDS surfactants at different compositions, at
			different interfacial areas of the simulation
			box ($A = L_x \times L_y$). Top figures, $L_x = L_y = 60$~\AA. Bottom
			figures $L_x = L_y = 120$~\AA. Pink color for the CTAB and blue for the SDS
			surfactants. Counterions are represented by dots and water is removed
			for better visualization of the monolayers.}
		\label{fig3}
	\end{center}
\end{figure}

 With the current surfactant LJ parameters, three monolayers were
prepared with mixtures of CTAB/SDS, CTAB/DTAB and CTAB/Betaine at
distinct compositions, 75/25, 50/50 and 25/75. Initial simulations
were conducted with the same box dimensions as described in the previous
section, i.e.,  $L_x = L_y = 5$~nm and $L_z = 15$~nm using 
an area per molecule of 26 \AA$^2$/molecule for the CTAB/SDS
and 40 \AA$^2$/molecule for the CTAB/DTAB and CTAB/Betaine mixtures.
The area per molecule is defined as the
interfacial area divided by the number of surfactants 
($L_x\times L_y$ / Num. surfactants). With these simulation boxes the
surface tension was calculated, although at these conditions
the results lead to large errors. Even negative values were found
for the asymmetric compositions, which have no physical
meaning (figure~\ref{fig2}a).
Exploring possible sources of errors,  from the snapshots
of the last configuration of the monolayers at small interfacial areas it is observed 
that they are distorted, i.e., they
are not planar (top of figure \ref{fig3}). Therefore, equation~\eqref{eq_3_1} cannot
be used, or in other words its application yields the not-physical results.
Interesting are the surfactant molecules in the bulk water phase forming
micelle-like structures for some surfactant
compositions (bottom of figure \ref{fig3}).

 We analyze the size effects on the surface tension data
by increasing the interfacial area of the box in different simulations.
Figure \ref{fig2} shows the
values of the surface tension as function of the box length,
by keeping the area per molecule constant. The simulations
were carried out for all the mixtures, for different compositions,
and in all cases similar trends were observed, i.e., the surface
tension fluctuates for small interfacial areas and it goes to constant values 
at large areas, for box lengths above $L_x = L_y = 120$~\AA.
It is worth mentioning that in 
these simulations, the number of surfactant molecules changes (to keep
the area per molecule fixed), and also changes the number of water molecules to have
a sufficient bulk water phase, at least 3 nm layer thick in the $z$-direction.
We also imposed $L{_z} = 3L{_x}$ and a cutoff radius of 2.5~nm, to have
reliable surface tension data as suggested in previous works \cite{13}.

\begin{figure}[h]
	%	\vspace{1cm}
	\begin{center}
		\includegraphics[width=3.in]{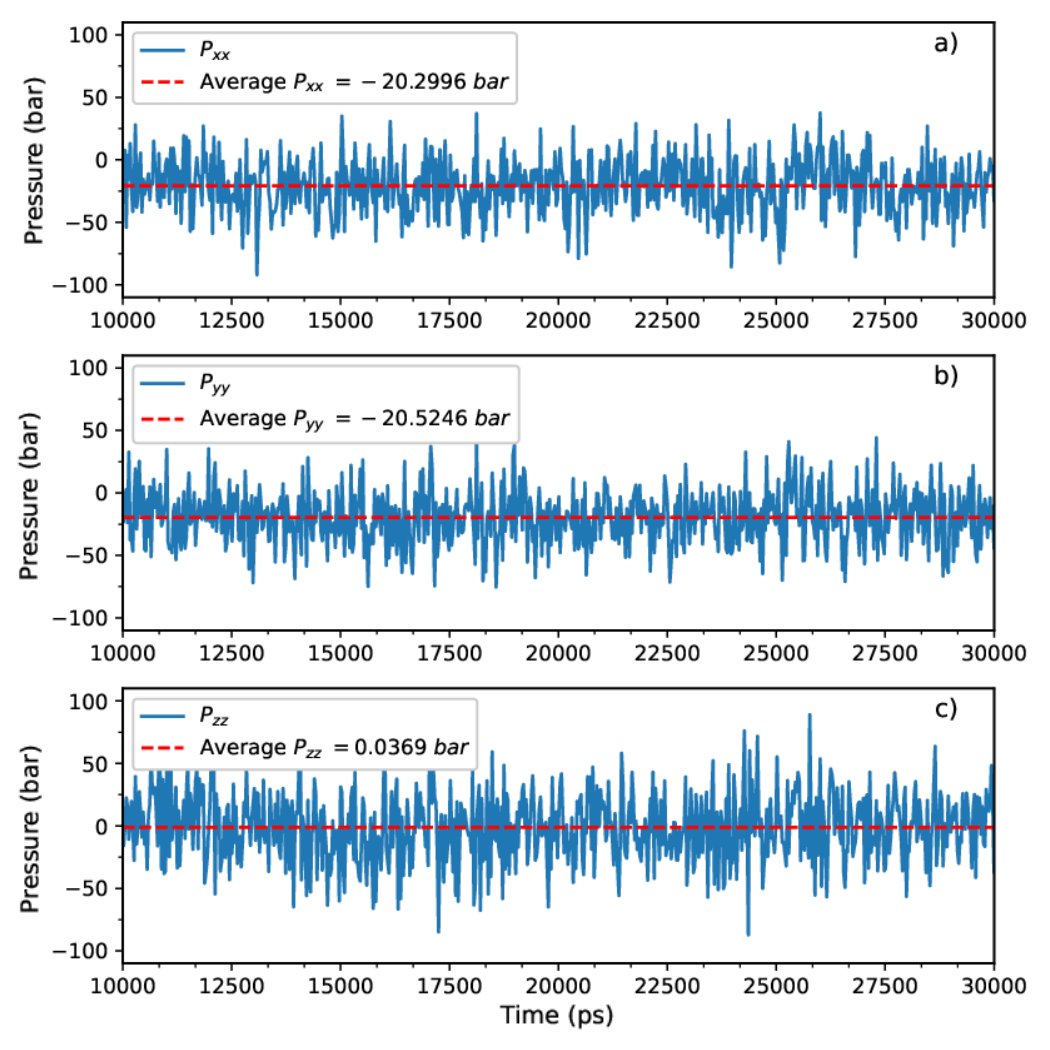}
		\caption{(Colour online) Components of the pressure tensor as function of
			the simulation time for the monolayer with the CTAB/SDS mixture,
			a) P$_{xx}$, b) P$_{yy}$, c) P$_{zz}$.}
		\label{fig4} 
	\end{center}
\end{figure}

 Simulations were run up to 80 ns and similar behaviour
was observed, the monolayers lost their planar structure for
small box lengths ($L_x$ and $L_y$). At large areas, they are flat
enough to consider that the equation~\eqref{eq_3_1} can be used correctly. Moreover,
for large box lengths, the components of the pressure tensor,
$P_{xx}$ and  $P_{yy}$ have similar values as expected.
It is important to mention that in the simulations, the pressure tensor
components show large fluctuations, although the average value
remains constant (figure~\ref{fig4}).

\begin{figure}[h]
	%	\vspace{1cm}
	\begin{center}
		\includegraphics[width=2.8in]{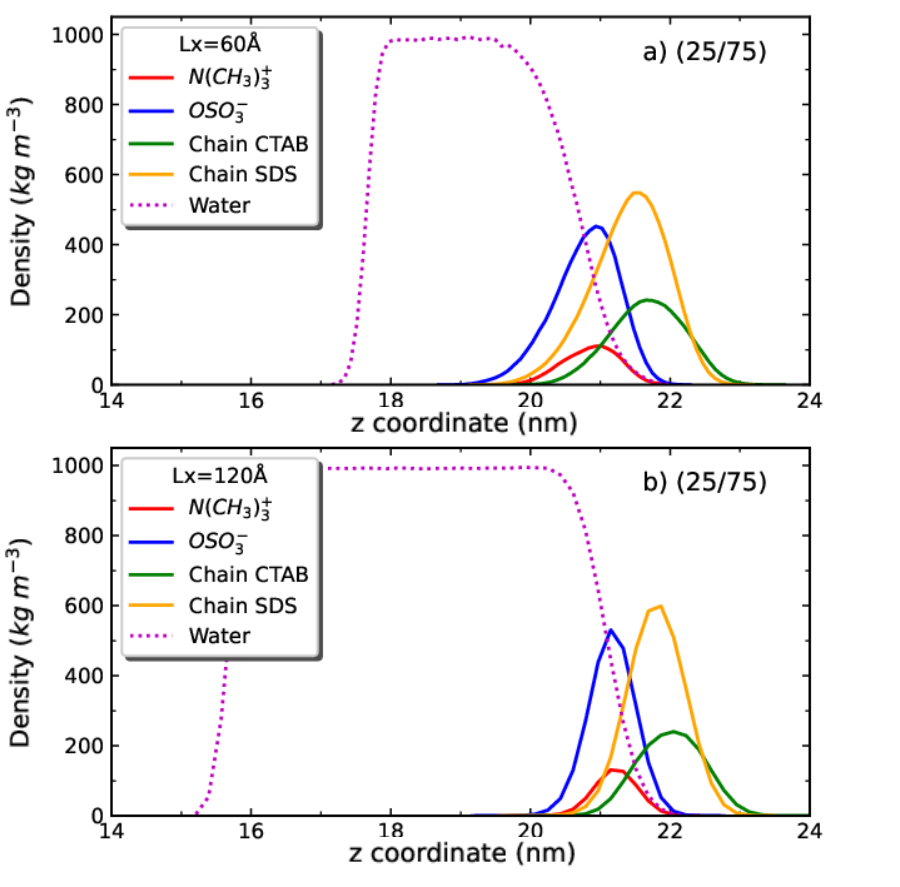}
		\caption{(Colour online) Mass density profiles for the CTAB/SDS monolayer at
			the 25/75 composition. Top: small box area. Bottom: large box area.}
		\label{fig5} 
	\end{center}
\end{figure}

 In order to better understand the behaviour of the monolayers at
the interface, in the next sections we study only the CTAB/SDS system more in detail, similar tendencies were observed for all the mixtures.
The analysis is carried out for two box lengths, in the fluctuating
region of figure \ref{fig2}, small area ($L_x = L_y = 60$~\AA) and
in the large area region ($L_x = L_y = 120$~\AA).

\begin{figure}[h]
	%	\vspace{1cm}
	\begin{center}
		\includegraphics[width=2.8in]{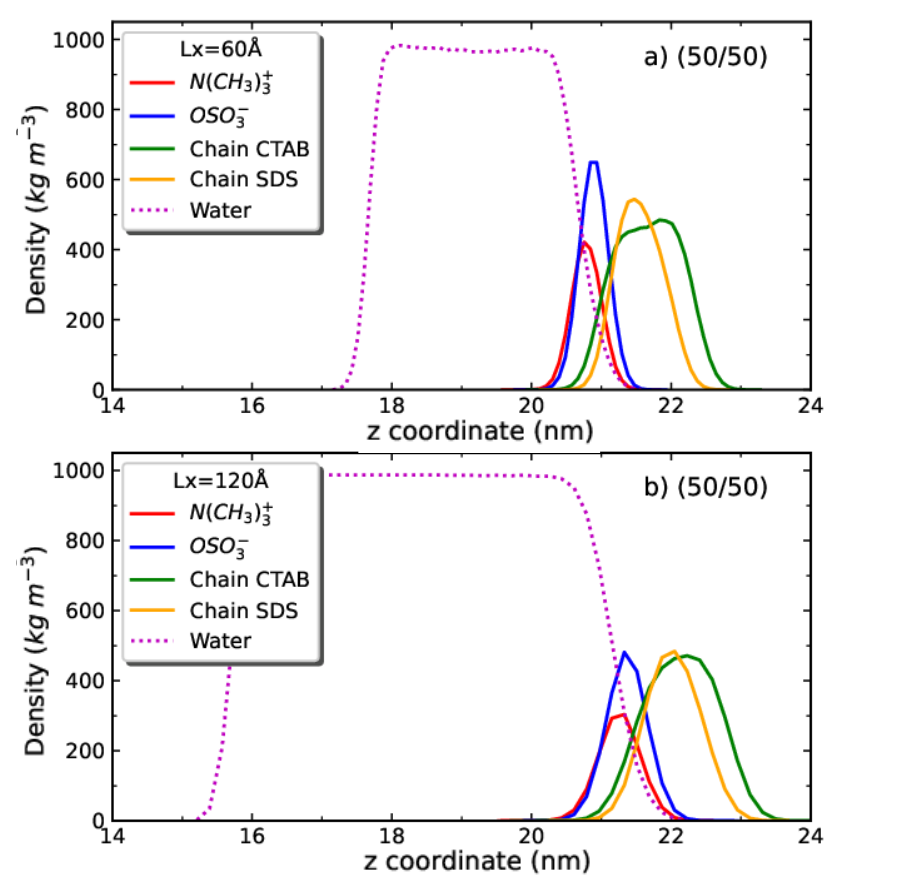}
		\caption{(Colour online) Mass density profiles for the CTAB/SDS monolayer at
			the 50/50 composition. Top: small box area. Bottom: large box area.}
		\label{fig6} 
	\end{center}
\end{figure}

\subsection{Density profiles for the monolayers of surfactant mixtures}

 Mass density profiles were calculated to determine the structure of
the monolayers at the interface, defined as
\begin{equation}
  \rho(z) = \frac{M(z)}{L_x \times L_y \times \Delta Z},
\end{equation}
 where $M(z)$ is the mass of the calculated species
in the volume $L_x \times L_y \times \Delta Z$. Since we are only interested
in the monolayers at the interface, the analysis of the CTAB/SDS mixture
at the 25/75 and 75/25 compositions, was done without considering the
micelles formed in the water phase at large
areas (see bottom of figure \ref{fig3}).

\begin{figure}[h]
	%	\vspace{1cm}
	\begin{center}
		\includegraphics[width=2.8in]{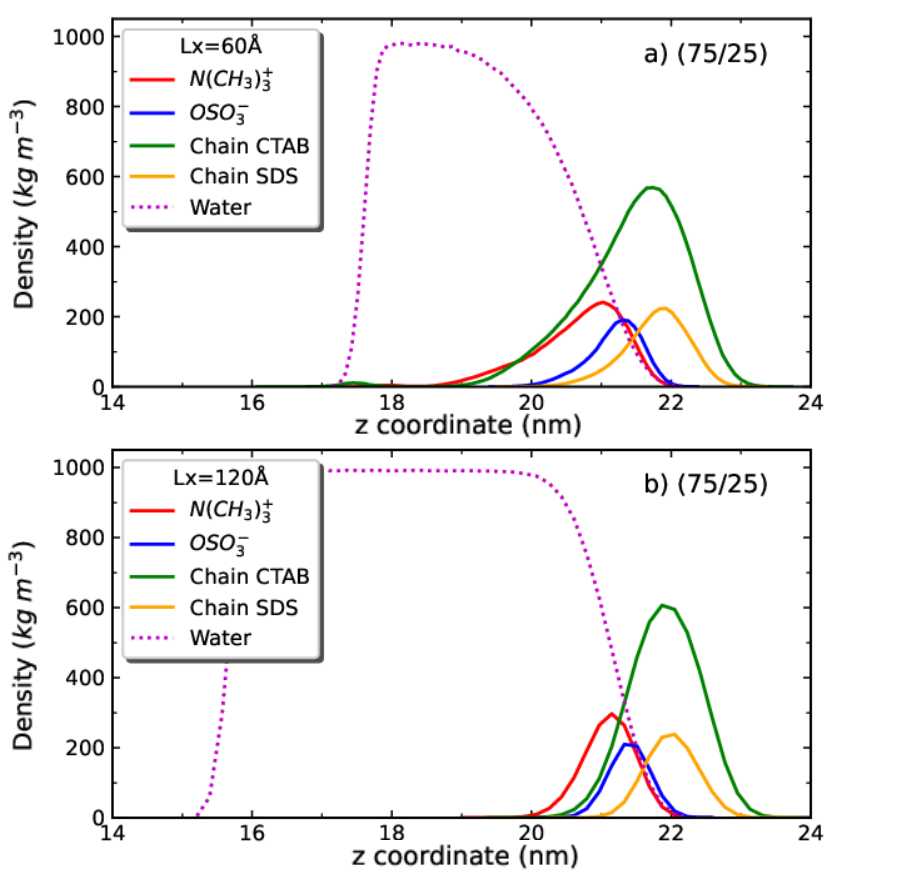}
		\caption{(Colour online) Mass density profiles for the CTAB/SDS monolayer at
			the 75/25 composition. Top: small box area. Bottom: large box area.}
		\label{fig7} 
	\end{center}
\end{figure}

 The results for the mass density profiles are shown in
figures \ref{fig5}--\ref{fig7}.
The headgroup profiles [OSO$^-_3$ for SDS and N(CH$_3$)$^+_3$ for CTAB]
are narrower for the systems with large box
lengths than those with small lengths indicating that the monolayers at
large interfacial areas are nearly planar. The monolayers in small areas,
are not planar and then equation~\eqref{eq_3_1} cannot be used
or incorrect values for the surface tension could be obtained.

\begin{figure}[h]
	%	\vspace{1cm}
	\begin{center}
		\includegraphics[width=3.in]{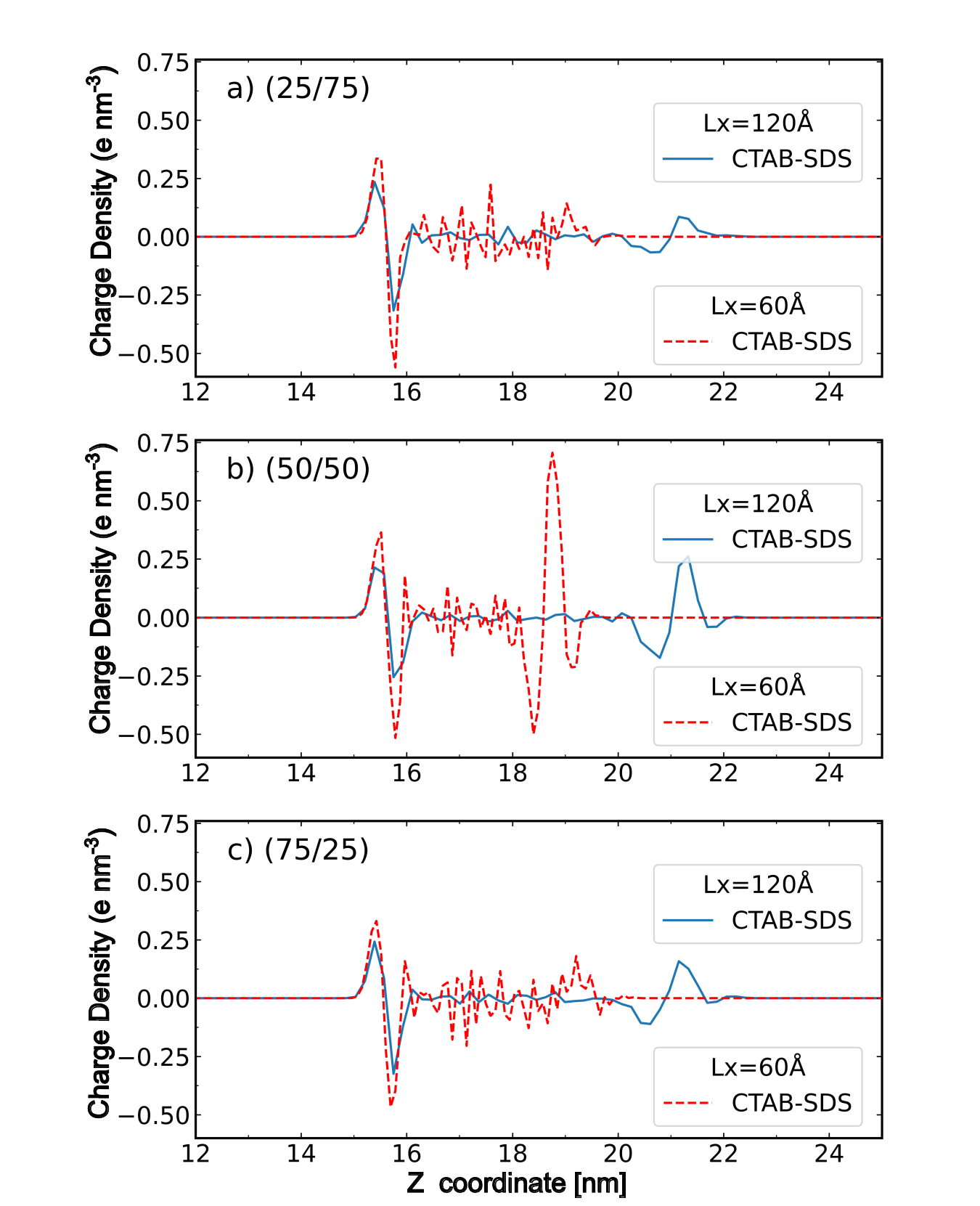}
		\caption{(Colour online) Charge density profiles for the CTAB/SDS monolayer
			at different surfactant compositions, a) 25/75, b) 50/50, c) 75/25,
			and for different box areas (box lengths $L_x = L_y$).}
		\label{fig8} 
	\end{center}
\end{figure}

 Charge distribution profiles can also give us information
about the distribution of the monolayers at the water/air interface.
These profiles are calculated as,
\begin{equation}
  \rho_q(z) = \frac{q(z)}{L_x \times L_y \times \Delta Z},
\end{equation}
 where $q(z)$ is the total charge in the volume
$L_x \times L_y \times \Delta Z$. 
The studies of the profiles were also carried out for the monolayers
simulated with small and large interfacial areas. In figure \ref{fig8} it
is possible to observe that the charge density profiles are nearly
zero, far from the interfaces (blue lines), for the systems simulated with
a large area ($L_x = L_y = 120$~\AA). For the monolayers simulated with the
small area ($L_x = L_y = 60$~\AA), an excess of charge in
the middle of the systems (red lines) is noted, most likely caused by the
surfactants headgroups. This last result suggests that not all
surfactants are located at the interface, i.e., planar monolayers
are not formed.

\subsection{Order of the surfactant tails and tilt angle}

\begin{table}[hb]
	\caption{\label{tab5} \small{Configurational energy	and the average order parameter, $\langle S_{zz} \rangle$, for the CTAB/SDS mixture at different compositions.}}
	\small
	\begin{center}
		%\vspace{.5cm}
		\begin{tabular}{|c|c|c|c|c|c|c|}
			\hline
			surfactant & Total energy & Total energy &
			$\langle S_{zz} \rangle$ & $\langle S_{zz} \rangle$
			& $\langle S_{zz} \rangle$ & $\langle S_{zz} \rangle$\\
			composition & $L_x=60$ \AA & $L_x=120$ \AA & $L_x=60$ \AA
			& $L_x=120$ \AA & $L_x=60$ \AA
			& $L_x=120$ \AA \\
			CTAB/SDS & Kj/mol/atom & Kj/mol/atom & CTAB & CTAB & SDS & SDS \\
			\hline
			25/75  & $-110$ & $-180$ & 0.433 & 0.432 & 0.252 & 0.274\\
			50/50  & $-100$ & $-150$ & 0.509 & 0.518 & 0.294 & 0.308\\
			75/25  &  $-93$ & $-170$ & 0.348 & 0.434 & 0.213 & 0.227 \\
			\hline
		\end{tabular}
	\end{center}
\end{table}

 Additional information about the structure of the monolayers
with small and large box areas can be obtained from an order parameter
which tells us how the surfactants are arranged within them.
In experiments, the ordering of the surfactant hydrocarbon tails
is characterized by the so-called deuterium order parameter, $S_{CD}$,
related to the average inclination of the C--H bond, in the CH$_n$
groups, with respect to the surface normal. However, in computer simulations
the order parameter is calculated, in the united atom model, by the following
equation \cite{20},
\begin{equation}
  S_{CD}  = \frac{2}{3} S_{xx} + \frac{1}{3} S_{yy},
\end{equation}
 with,
\begin{equation}
  S_{ij}  = \frac{1}{2} \langle 3\cos\theta_i \cos\theta_j - \delta_{ij}
    \rangle ,
\end{equation}
 where $i,j =x,\,y,\,z $ and $\theta_i$ is the angle between the
$i$-th molecular
axis and the normal to the interface. In this work we 
calculated the $S_{zz}$ order parameter. It is worth mentioning that
$S_{zz} = -0.5$ correspond to complete order parallel to the interface whereas
$S_{zz} = 1.0$ is complete order in the direction normal to the interface.
However, here it is more convenient to
use the average of the order parameter, $\langle S_{zz} \rangle$,
to study the order in the tails \cite{hec}.
Then, the analysis of the order parameter was carried out over the carbons
that are at the water/air interface. In table \ref{tab5}
the $\langle S_{zz} \rangle$ values for the first four carbons
are given for both surfactants, CTAB and SDS, for a small
and for a large interfacial area. It is noted that the tails
are more ordered for large areas than for small ones, suggesting
that the surfactants accommodate better in the monolayer at the large
interfacial area.

\begin{table}[h]
	\caption{\label{tab6} \small{Tilt angle ($\theta_t$) of the surfactant chains for the CTAB/SDS mixture at different compositions.}}
	\small
	\begin{center}
		%\vspace{.5cm}
		\begin{tabular}{|c|c|c|c|c|}
			\hline
			surfactant & $\theta_t$ & $\theta_t$ & $\theta_t$ & $\theta_t$\\
			composition & $L_x=60$ \AA & $L_x=120$ \AA & $L_x=60$ \AA & $L_x=120$ \AA \\
			CTAB/SDS & CTAB & CTAB & SDS & SDS \\
			\hline
			25/75 & 44$^0$ & 51$^0$ & 49$^0$ & 50$^0$ \\
			50/50 & 39$^0$ & 41$^0$ & 47$^0$ & 47$^0$ \\ 
			75/25 & 49$^0$ & 49$^0$ & 54$^0$ & 55$^0$ \\
			\hline
		\end{tabular}
	\end{center}
\end{table}

 The structure of the monolayers can also be analyzed by the tilt
angle of the surfactant chains. The angle is measured as,
$\theta_t$ = $\delta_z$/$\delta_t$, where $\delta_z$ is the
average projection of the chains along the normal to the interface and
$\delta_t$ is the total length of the tails (from the last to the first
carbon atom). The results are shown in table \ref{tab6}. It is observed that the
tilt angle in general is a bit lower for the mixtures in the small
area than in the large one, regardless of the surfactant, i.e.,
the tails are straighter with respect to the interface at the large box 
area allowing better arrangement of the surfactants in the
monolayers, in agreement with the order
parameter $\langle S_{zz} \rangle$ results.

\begin{figure}[h]
	%	\vspace{1cm}
	\begin{center}
		\includegraphics[width=3.2in]{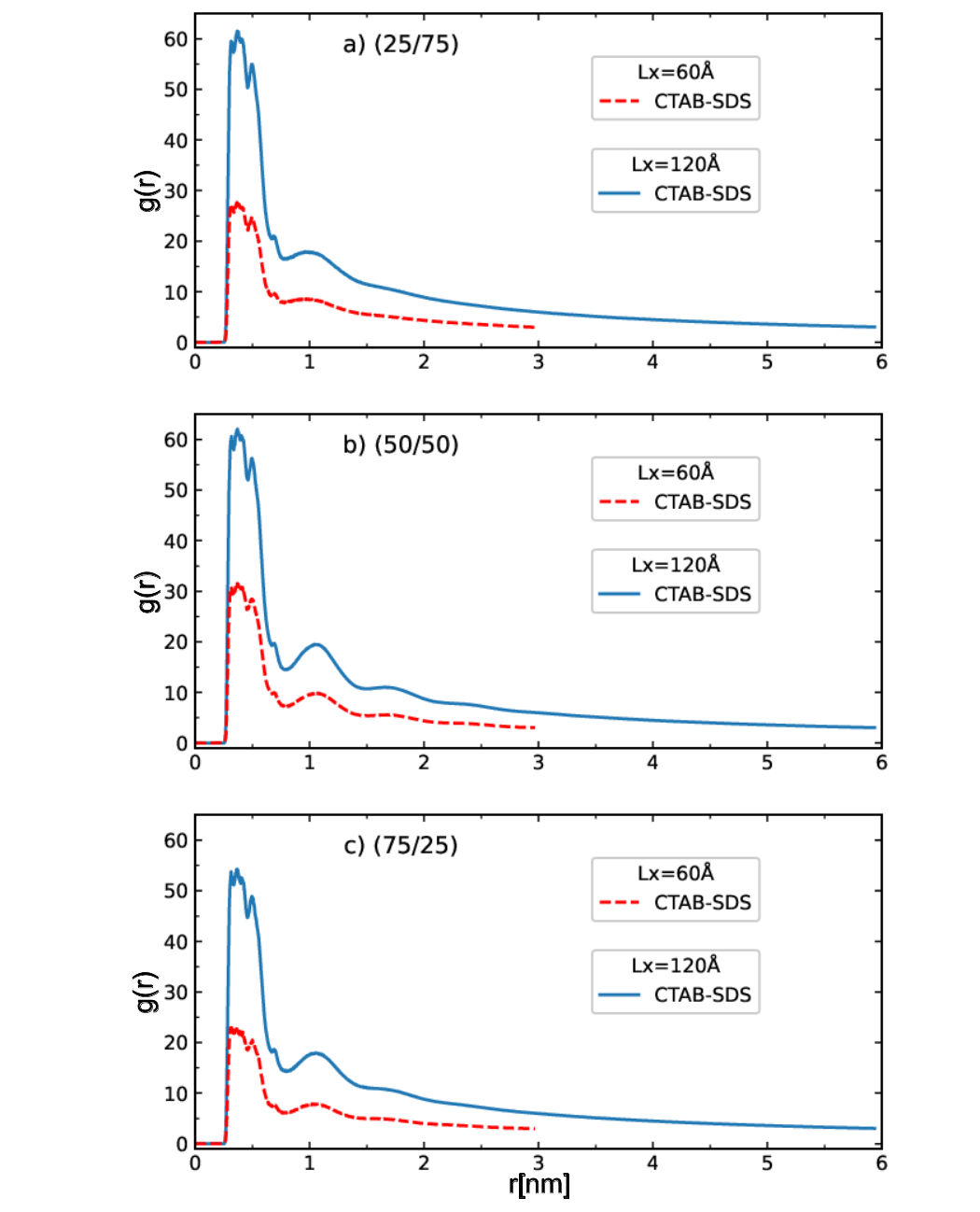}
		\caption{(Colour online) Radial distribution functions, $g(r)$, of the
			surfactant headgroups for the CTAB/SDS
			monolayer at different compositions a) 25/75, b) 50/50, c) 75/25,
			and for different box areas (box lengths $L_x = L_y$).}
		\label{fig9} 
	\end{center}
\end{figure}

 The arrangement between the different surfactants inside
the monolayer was also studied in terms of pair distribution
functions, $g(r)$. The $g(r)$ were calculated between the headgroups
of each surfactant and the results are given in figure \ref{fig9}.
For all the CTAB/SDS compositions, the peaks in the $g(r)$ are higher
for the systems with the large area (length of $L_x = L_y = 120$~\AA)
than for the small area ($L_x = L_y = 60$~\AA). These results
suggest that there are more surfactants close to each other
in the monolayers simulated with large areas than with small ones and
consequently, the monolayers are more ``compact'' in the first case.

 One last analysis to compare the monolayers for small and
large interfacial areas can be given in terms of the configuration energies,
calculated over the monolayers only,
and the results are shown in table \ref{tab5}. Energies are negatively
greater for the
monolayers with large areas than for the small areas,
indicating that the systems in large areas are more stable.
The values of energies  are divided by the total number of atoms in the monolayer.

\section{Conclusions}

 The mixtures of  surfactants are
of great relevance to investigate in several industrial and technological
applications (foaming, detergency, etc.), where the surface tension is used as a
parameter to determine their synergistic effects \cite{ros, ros1}.
In the present paper, we carried out 
molecular dynamics of monolayers with ionic and non-ionic surfactants
to study the surface tension at the water/air interface.
We pointed out the importance of choosing right simulation boxes 
to obtain reliable surface tension values of
surfactant monolayers. Unsuitable conditions, such as calculations
with small interfacial areas, can lead to artificial effects
in the surface tension values. For instance, fluctuations in
the surface tension data are observed due to the periodic boundary
conditions and the finite box size. In the case of simulations of surfactant
mixtures with small areas, the surface tension calculations
can give non-physical values such as negative ones.
It was found that reasonable values of the surface tension are obtained
when the simulations are conducted with large box areas (box lengths
of above 120 \AA), with cutoffs of 2.5 nm and using a sufficient number of water
molecules to form bulk phases at least 30 \AA {} thick.

 Several interfacial problems, e.g., the description of the
solid-liquid interface requires the knowledge of the surface tension.
Then, the wetting
properties of such complex interfaces can be studied. Therefore, we expect
that our results may be useful and stimulate theoretical developments
along lines of research discussed in \cite{ors1, ors2, ors3}.

\section*{Acknowledgements}
 This work was supported by DGAPA-UNAM-Mexico grant PAPIIT-IG101621,
Conacyt-Mexico grant A1-S-29587 and DGTIC-UNAM LANCAD-UNAM-DGTIC-238
for supercomputer facilities.
We also acknowledge Alejandro-Pompa and Cain Gonzalez-Sanchez
for technical support in the computer calculations.

%\newpage
%\vspace{20cm}

\newpage

\ukrainianpart

\title{Ефекти скінченого розміру та оптимізація розрахунку поверхневого натягу в сумішах поверхнево-активних речовин на межі розділу ``рідина-пара''}
\author{С. Х. Гузман-Валенсія\refaddr{label1}, Х. Торіс-Салінас\refaddr{label1},
	Е. Еспіноса-Хіменес\refaddr{label1}, А.~Б.~Салазар-Арріага\refaddr{label1},
	Х. Л. Лопес-Сервантес\refaddr{label2}, 
	Е. Домінгес\refaddr{label1}}
\addresses{
	\addr{label1}	Інститут матеріалознавства, Національний автономний університет Мехіко, UNAM Cd. 04510, Мехіко, Мексика
	\addr{label2} Хімічний факультет, Національний автономний університет Мехіко, UNAM Cd. 04510, Мехіко, Мексика
}

\makeukrtitle

\begin{abstract}
	\tolerance=3000%
	Досліджено поверхневий натяг моношарів із сумішами аніонних та неіонних ПАР на межі розділу ``рідина-пара''. У попередніх роботах було виявлено, що розрахунки поверхневого натягу простих плинів підтверджують штучні коливання для малих міжфазних ділянок. Це вказує на те, що дані поверхневого натягу міняються через ефект скінченого розміру та періодичні граничні умови. У випадку моделювання моношарів, що складаються із сумішей ПАР, поверхневий натяг не тільки міняється на невеликих ділянках, але також може давати нефізичні значення, зокрема, бути від'ємним. Аналіз моношарів із різними сумішами ПАР, іонними (DTAB, CTAB, SDS) та неіонними (SB3-12), проводився для профілів густини, параметрів порядку та парних кореляційних функцій для малих і великих комірок моделювання, і всі вони демонструють подібність поведінки. Флуктуації та нефізичні значення поверхневого натягу корегуються, коли розглядаються комірки з великими міжфазними площами. Результати показують, що для отримання надійних значень поверхневого натягу у комп’ютерному експерименті важливо вибрати не лише правильне силове поле, але й адекватний розмір комірки моделювання.
	
	\keywords ефекти скінченого розміру, молекулярна динаміка, поверхневий натяг, моношаровi сумiшi ПАР, межа роздiлу``вода-повiтря''
	
\end{abstract}

\lastpage
\end{document}